\documentclass{elsart}
\usepackage{amssymb}
\usepackage{graphics}

\begin{document}

\begin{frontmatter}

\title{Field intensity distributions and polarization orientations in a vacuum-clad subwavelength-diameter
 optical fiber}

\author[a]{Fam Le Kien\corauthref{*}\thanksref{kien}},
\corauth[*]{Corresponding author. Tel.: +81-424-43-5476; Fax: +81-424-43-5507.}
\thanks[kien]{Also at Institute of Physics and Electronics, Vietnamese Academy of Science and Technology, Hanoi, Vietnam.}
\ead{fam@kiji.pc.uec.ac.jp}
\author[a]{J. Q. Liang},
\author[a]{K. Hakuta},
\author[a,b]{V. I. Balykin}

\address[a]{Department of Applied Physics and Chemistry, 
University of Electro-Communications, Chofu, Tokyo 182-8585, Japan}
\address[b]{Institute of Spectroscopy, Troitsk, Moscow Region, 142092, Russia}

\begin{abstract}
We study the properties of the field in the fundamental mode HE$_{11}$  of  a vacuum-clad 
\textit{subwavelength-diameter} optical fiber using the exact solutions of Maxwell's equations. 
We obtain simple analytical  expressions for the total intensity of the electric field. We discuss the origin of the deviations of the exact fundamental mode HE$_{11}$ from the approximate mode LP$_{01}$. 
We show that the thin thickness of the fiber and the high  contrast between the refractive indices of the silica core and the vacuum clad substantially modify the intensity distributions and the polarization properties of the field and its components, especially in the  vicinity of the fiber surface. One of the promising applications of  the field around the subwavelength-diameter fiber is  trapping and guiding of atoms by the optical force of the evanescent field.
\end{abstract}

\begin{keyword}
Subwavelength-diameter optical fiber;  Fundamental mode; Field intensity distribution; Polarization orientation; Atom trapping and guiding
\PACS 42.81.-i \sep 42.81.Dp \sep 42.81.Gs \sep 42.81.Qb
\end{keyword}

\end{frontmatter}

\section{Introduction}

Optical fibers with diameters from micrometers to millimeters have found many important applications
in  technology. The guiding properties of such fibers have been well studied \cite{fiber books}. 
Most fibers use  core and cladding materials with nearly the same refractive indices. Such fibers
are referred to as weakly guiding fibers. The description of weakly guiding fibers is greatly simplified due to the use of the linearly polarized (LP) modes, which are  approximate solutions of  Maxwell's  equations and are superpositions of two corresponding nearly degenerate eigenmodes.

Recently, thin waveguides have become attractive for a wide range of potential practical applications.
It has been pointed out that the intense evanescent field in the vicinity of a tapered fiber can be used as an atomic mirror \cite{Bures and Ghosh}. Generation of light with a supercontinuum spectrum in a thin tapered fiber has been demonstrated \cite{Birks}. 
The evanescent waves from zero-mode metal-clad subwavelength-diameter waveguides have been used for optical observations of single-molecule dynamics \cite{zero mode}. Several types of dielectric submicrometer- and nanometer-diameter wire waveguides have been fabricated, and their guiding properties have been 
investigated \cite{Mazur's Nature}. 
It has been proposed to use the optical force of an evanescent wave around a thin fiber for atom trapping and guiding \cite{our paper}. Thin fiber structures can be used as building blocks in the future micro- and nano-photonic devices.

Unlike the case of optical fibers with diameter larger than the wavelength, many properties of 
subwavelength-diameter fibers have not been adequately investigated. 
In such a thin fiber, the original silica core is vanishing. Therefore, the original silica clad acts like a core while  the surrounding vacuum (or air) acts like a clad.
Due to the high refractive-index contrast between  the silica core and the vacuum clad, the description of subwavelength-diameter fibers in terms of weakly guided LP modes is questionable. 
It has been demonstrated that such fibers have interesting properties such as 
high power density at the fiber surface and cylindrical asymmetry in the field distribution \cite{Bures and Ghosh}, large penetration length of the evanescent wave \cite{our paper}, enhancement  of the 
power fraction of the field outside the fiber  \cite{our paper,Mazur's OpEx}, and  large
waveguide dispersion  \cite{Mazur's OpEx}. 
However, the intensity distributions and polarization orientations
of the fields in subwavelength-diameter fibers have not been studied.

In this paper we  study the field intensity distributions and polarization orientations in the fundamental mode HE$_{11}$  of  a vacuum-clad \textit{subwavelength-diameter} optical fiber using the \textit{exact} solutions of Maxwell's equations. We show that the thin thickness of the fiber and the high  contrast between the refractive indices of the silica core and the vacuum clad substantially modify the intensity distributions and the polarization orientations of the field and its components. The study of such properties is necessary not only for academic interest but also for practical applications such as  trapping of atoms by the optical force of an evanescent wave around a thin fiber \cite{our paper}. 

Before we proceed, we note that, in related problems, the decay of an atom in the presence of a fiber
was considered for the first time by Katsenelenbaum in 1949 \cite{Katsen}. Excitation of fiber modes with
a dipole source and influence of a fiber on the decay rate of a single atom were investigated in 
\cite{Katsen,Jhe,Jak,Klimov}. In the present paper, we do not consider how the modes can be excited.
Due to the adiabatic tapping condition \cite{taper}, the fundamental mode HE$_{11}$ of  
a subwavelength-diameter fiber can be excited by the standard coupling techniques \cite{Mazur's Nature}.

The paper is organized as follows. In Sec.\ \ref{sec:linear} 
we describe the general model of a thin fiber and examine a fundamental mode with  quasi-linear polarization.
In Sec.\ \ref{sec:circular} we study a fundamental mode  with  rotating (circulating) polarization.
Our conclusions are given in Sec.~\ref{sec:summary}.

\section{Thin fiber and fundamental mode with  quasi-linear polarization}
\label{sec:linear}

Consider a thin  single-mode optical fiber that has a cylindrical silica core of radius $a$ and refractive index $n_1$ and  an infinite vacuum clad
of refractive index $n_2=1$.
Such a fiber can be prepared using taper fiber technology. 
The essence of the technology is to heat and 
pull a single-mode optical fiber to a very thin thickness 
maintaining the taper condition to keep adiabatically 
the single-mode condition \cite{Birks,taper}. 
Due to tapering, the original core  is almost vanishing.  
Therefore, the refractive
indices that determine the guiding properties of  the tapered fiber are the refractive index of the original silica clad and the refractive index of the surrounding vacuum. 
The refractive index and the radius of the tapered silica clad will be henceforth referred to simply as the fiber refractive index $n_1$ and the fiber radius $a$, respectively. 

We send a light of wavelength $\lambda$,  frequency $\omega$, and free-space wave number
$k=2\pi/\lambda=\omega/c$ through the fiber.
Under the condition $V\equiv ka\sqrt{n_1^2-n_2^2}<V_c\cong2.405$,
the fiber can support only one mode, referred to as the fundamental mode HE$_{11}$. 
The longitudinal propagation constant $\beta$ of this mode is determined by the eigenvalue equation \cite{fiber books}
\begin{eqnarray}
\frac{J_0(h a)}{h a J_1(h a)}&=&
-\frac{n_1^2+n_2^2}{2n_1^2}\frac{K_1'(q a)}{q a K_1(q a)}+ \frac{1}{h^2 a^2}
\nonumber\\&&\mbox{}
-\Bigg\{\left[\frac{n_1^2-n_2^2}{2n_1^2}\frac{K_1'(q a)}{q a K_1(q a)}\right]^2
+\frac{\beta^2}{n_1^2 k^2}\left(\frac{1}{q^2a^2}+\frac{1}{h^2a^2}\right)^2 \Bigg\}^{1/2}.
\label{0}
\end{eqnarray}
Here the parameters $h=(n_1^2k^2-\beta^2)^{1/2}$ and $q=(\beta^2-n_2^2k^2)^{1/2}$ characterize the fields inside and outside the fiber. The notation $J_n$ and $K_n$ stand for the  Bessel functions of the first kind and the modified Bessel functions of the second kind, respectively.

We first study a fundamental mode with  quasi-linear  polarization.
In the cylindrical coordinates ($r,\varphi,z$), the solutions of  Maxwell's equations for 
the Cartesian components of the electric field $\mathbf{E}$ in such a mode are given, 
for  $r<a$ (inside the fiber), by \cite{fiber books}
\begin{eqnarray}
E_x&=&-iA\frac{\beta}{2h}[(1-s)J_0(hr)\cos\varphi_0
-(1+s)J_2(hr)\cos(2\varphi-\varphi_0)]\,e^{i(\omega t-\beta z)},
\nonumber\\
E_y&=&-iA\frac{\beta}{2h}[(1-s)J_0(hr)\sin\varphi_0
-(1+s)J_2(hr)\sin(2\varphi-\varphi_0)]\,e^{i(\omega t-\beta z)},
\nonumber\\
E_z&=& AJ_1(hr)\cos(\varphi-\varphi_0)\,e^{i(\omega t-\beta z)},
\label{5}
\end{eqnarray}
and, for   $r>a$ (outside the fiber), by \cite{fiber books}
\begin{eqnarray}
E_x&=&-iA\frac{\beta}{2q}\frac{J_1(ha)}{K_1(qa)}[(1-s)K_0(qr)\cos\varphi_0
\nonumber\\
&&\mbox{}+(1+s)K_2(qr)\cos(2\varphi-\varphi_0)]\,e^{i(\omega t-\beta z)},
\nonumber\\
E_y&=&-iA\frac{\beta}{2q}\frac{J_1(ha)}{K_1(qa)}[(1-s)K_0(qr)\sin\varphi_0
\nonumber\\
&&\mbox{}+(1+s)K_2(qr)\sin(2\varphi-\varphi_0)]\,e^{i(\omega t-\beta z)},
\nonumber\\
E_z&=& A\frac{J_1(ha)}{K_1(qa)}K_1(qr)\cos(\varphi-\varphi_0)\,e^{i(\omega t-\beta z)},
\label{7}
\end{eqnarray}
where $s=[(qa)^{-2}+(ha)^{-2}]/[{J_1'(ha)}/{haJ_1(ha)}+{K_1'(qa)}/{qaK_1(qa)}]$.
The coefficient $A$ is determined by the normalization condition.
The angle $\varphi_0$ determines the orientation axis of the polarization  of the field.
The two  sets of solutions corresponding to $\varphi_0=0$ and $\varphi_0=\pi/2$ express two different polarizations,
aligned along the $x$ and $y$ axes, respectively.

For practical applications such as  trapping of atoms by the optical force of an evanescent wave around a thin fiber \cite{our paper}, it is necessary to calculate the optical potential, which is proportional to the total intensity $|E|^2$ of the electric field. A  rigorous expression for $|E|^2$ can be easily obtained 
with the help of Eqs.~(\ref{5}) and (\ref{7}). 
For $r<a$, we find
\begin{eqnarray}
|E|^2&=&g_{\mathrm{in}}\{J_0^2(hr)+u J_1^2(hr)+fJ_2^2(hr)
\nonumber\\&&\mbox{}
+[u J_1^2(hr)
-f_pJ_0(hr)J_2(hr)]\cos[2(\varphi-\varphi_0)]\}.
\label{12}
\end{eqnarray}
For $r>a$, we get
\begin{eqnarray}
|E|^2&=&g_{\mathrm{out}}\{K_0^2(qr)+w K_1^2(qr)
+fK_2^2(qr)
\nonumber\\&&\mbox{}
+ [w K_1^2(qr)+f_pK_0(qr)K_2(qr)]\cos[2(\varphi-\varphi_0)]\}.
\label{11}
\end{eqnarray}
Here we have introduced the parameters
$u= 2h^2/\beta^2(1-s)^2$,
$w= 2q^2/\beta^2(1-s)^2$,
$f=(1+s)^2/(1-s)^2$, and $f_p=2(1+s)/(1-s)$.  
We have also introduced the notation
$g_{\mathrm{in}}=|A|^2/2u$ and 
$g_{\mathrm{out}}=|A|^2J_1^2(ha)/2wK_1^2(qa)$.
The terms $J_0^2(hr)$ and  $K_0^2(qr)$ in the expressions (\ref{12}) and (\ref{11}), respectively, correspond to the total intensity of the electric field in the mode LP$_{01}$. 
The other terms  describe the deviations of the exact mode HE$_{11}$ from the approximate  mode LP$_{01}$.

The above expressions  are mathematically valid for any core radius $a$ and any pair of refractive indices $n_1$ and $n_2$ where $n_1>n_2$. Before applying these expressions to thin tapered  fibers, we  recall
the case of conventional single-mode fibers, where $\Delta\equiv (n_1-n_2)/n_1\ll1$ and $a>\lambda$. 
In this case, we have \cite{fiber books}
$s=-1+(qa/V)^2[haJ_0(ha)/J_1(ha)]\Delta+O(\Delta^2)\cong -1$.
Consequently, the $\varphi$-dependent terms containing  $(1+s)J_2(hr)$ and $(1+s)K_2(qr)$
in the expressions (\ref{5}) and (\ref{7}) are negligible. Hence, the transverse component $E_y$ or $E_x$ will be zero if $\varphi_0=0$ or $\varphi_0=\pi/2$, respectively. 
In addition, we have $h,q\ll\beta$, so
the longitudinal component $E_z$  
is small.  
Thus, an exact HE$_{11}$ mode with quasi-linear polarization  of a conventional weakly guiding fiber can be approximated by  a LP$_{01}$ mode \cite{fiber books}. 
The electric field in the LP$_{01}$ mode is given by 
$\mathbf{E}=F
\hat{\mathbf{e}}\, e^{i(\omega t-\beta z)}$,
where
$F= {\mathcal A} J_0(hr)$ for   $r<a$ and 
$F={\mathcal A} [J_0(ha)/K_0(qa)] K_0(qr)$ for $r>a$.
The polarization vector $\hat{\mathbf{e}}$ can be chosen as $\hat{\mathbf{e}}=\hat{\mathbf{x}}$ or $\hat{\mathbf{e}}=\hat{\mathbf{y}}$.
The intensity distribution of the LP$_{01}$ mode is cylindrically symmetric.

\begin{figure}
\begin{center}
  \includegraphics{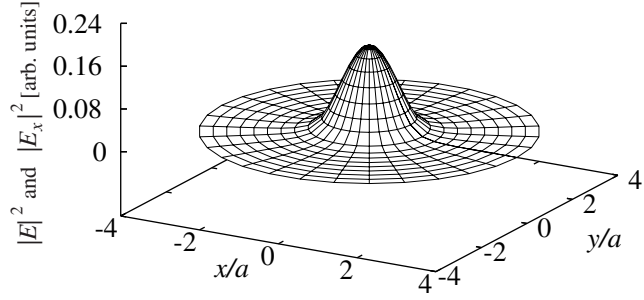}
 \end{center}
\caption{Total intensity $|E|^2$ and $x$-component
intensity $|E_x|^2$ of the electric field in an exact fundamental mode HE$_{11}$ with a quasi-linear polarization, along the $x$ direction, for the parameters  of a conventional weakly guiding fiber.
The two profiles are indistinguishable because
the intensities of the $y$ and $z$ components of the field are negligible.
The parameters used: $a=4$ $\mu$m, $\lambda=1.3$ $\mu$m, $n_1=1.4469$, $n_2=1.4419$, and $\varphi_0=0$.
}
\label{fig1}
\end{figure}

In Fig.~\ref{fig1}, we plot the total intensity $|E|^2$  and the $x$-component
intensity $|E_x|^2$ of the electric field in an exact fundamental mode HE$_{11}$ with a quasi-linear polarization, along the $x$ direction, for the parameters of a conventional weakly guiding fiber. 
As seen, the transverse profiles of $|E|^2$ and  $|E_x|^2$ are indistinguishable. 
Thus the polarization of the field is almost completely linear.
We also observe that the two profiles are almost perfectly cylindrically symmetric.

We now study the case of vacuum-clad \textit{subwavelength-diameter} fibers, where the relations $\Delta\ll1$ and  $a>\lambda$ are not satisfied. In this case, the factor $1+s$ is not negligible.
More importantly, the decay parameter $qa$ of the evanescent wave may become sufficiently small that $K_1(qa)$ and $K_2(qa)$ are much larger than $K_0(qa)$. Hence, the terms containing $K_1(qr)$ and $K_2(qr)$ in the expressions (\ref{7})  may become significant in the outer vicinity of the fiber surface. 
Due to the factor $K_1(qr)$, the longitudinal component $E_z$ outside the fiber  may become substantial. Due to the factor $K_2(qr)$, the terms containing 
the trigonometric functions $\cos(2\varphi-\varphi_0)$ and $\sin(2\varphi-\varphi_0)$ in the expressions (\ref{7}) may become significant. These terms lead to the azimuthal dependences of the
transverse components $(E_x,E_y)$  of the field outside the fiber. 
In addition, when $qa$ is small and $n_1$ is much different from $n_2$,  
the parameter $h$ may become comparable to (or even larger than)
the parameter $\beta$. Due to this fact, the component $E_z$  of the field inside the fiber may also be
not negligible compared to the components $(E_x,E_y)$, respectively, see Eqs.~(\ref{5}). Thus the longitudinal component $E_z$  may be substantial in both  regions $r<a$ and $r>a$. So we see that the properties of the exact fundamental mode HE$_{11}$ may become substantially different from the linearly polarized mode LP$_{01}$.

To demonstrate the features of  vacuum-clad \textit{subwavelength-diameter} fibers, we perform numerical calculations. The single-mode condition for vacuum-clad silica-core fibers requires that $a/\lambda<0.36$, that is, the fiber radius be about three times smaller than the light wavelength.
We consider the case where  the fiber radius $a$ is small compared not only to the light wavelength but also to the  evanescent-wave penetration length $\Lambda=1/q$. 
Such  thin fibers can be used to trap atoms. Indeed, it has been shown  
that the optical force of a red-detuned evanescent wave around an optical fiber can balance
the centrifugal force under the condition $qa=a/\Lambda<0.93$ and that this condition can be achieved 
for  vacuum-clad silica-core fibers with  $a/\lambda<0.28$ \cite{our paper}. We study such thin fibers in the rest of this paper.

To be specific, we choose the fiber  radius $a=0.2$ $\mu$m  and the light wavelength $\lambda= 1.3$ $\mu$m for  simulations.
The corresponding refractive indices of the silica core and the vacuum clad are $n_1\cong 1.4469$ and $n_2=1$, respectively. 
The solution of the exact eigenvalue equation (\ref{0}) yields $ha\cong1.0075$, $qa\cong0.0827$, and  $\beta a\cong0.9702$.  The normalized size parameter is $V=ka\sqrt{n_1^2-n_2^2}
\cong 1.011<2.405$, indicating that the considered fiber is a single-mode fiber.
For the above parameters, we find $s\cong-0.9937$. 
Since the fiber radius $a$ is thin compared to the light wavelength $\lambda$, the penetration length 
of the evanescent wave is $\Lambda\cong12\,a$, large compared to the fiber radius \cite{our paper}. Because of this,  a majority of the field intensity distribution is in the  outside of the fiber \cite{our paper,Mazur's OpEx}.

\begin{figure}
\begin{center}
  \includegraphics{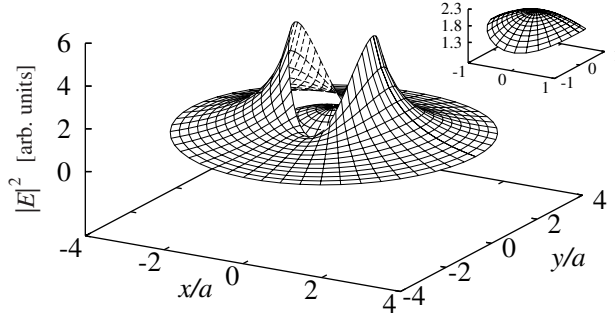}
 \end{center}
\caption{Cross-section profile of the total intensity $|E|^2$ of the electric field in a fundamental mode with  quasi-linear polarization. The inset shows the inner part of the profile, which corresponds to the field inside  the fiber. 
The parameters used: $a=0.2$ $\mu$m, $\lambda=1.3$ $\mu$m, $n_1=1.4469$, $n_2=1$, and $\varphi_0=0$.
}
\label{fig2}
\end{figure}

In Fig.~\ref{fig2}, we plot the total  intensity $|E|^2$ of the  electric field
in a fundamental mode with  quasi-linear polarization.  
We choose the $x$ axis as the major orientation axis of polarization ($\varphi_0=0$). 
As seen, the behaviors of the profiles inside and outside the fiber are very different from each other.
A conspicuous discontinuity of  the field intensity is observed at the fiber surface.
This discontinuity is due to the boundary condition for the normal (radial) component $E_r$ of the electric field. 
The high contrast between the refractive indices $n_1$ and $n_2$ of the silica core 
and the vacuum clad, respectively, makes this effect dramatic. 

As seen from Fig.~\ref{fig2}, the spatial distribution of the field intensity is not cylindrically symmetric at all.
A strong dependence of the field intensity on the azimuthal angle is observed in the outer vicinity
of the fiber surface. This dependence is due to
the terms $w K_1^2(qr)\cos[2(\varphi-\varphi_0)]$ and $f_p K_0(qr)K_2(qr)\cos[2(\varphi-\varphi_0)]$
in Eq.~(\ref{11}). 
These terms are substantial because  the small magnitude of the parameter $qa\cong0.0827$
makes $K_1(qr)$ and $K_2(qr)$ dominant over $K_0(qr)$ in the vicinity of the fiber surface. Indeed, we have
$K_1(qa)/K_0(qa)\cong 4.6$ and  $K_2(qa)/K_0(qa)\cong 111.7$. 
In the inner vicinity of the fiber surface, the azimuthal dependence of the field intensity is moderate. 

\begin{figure}
\begin{center}
  \includegraphics{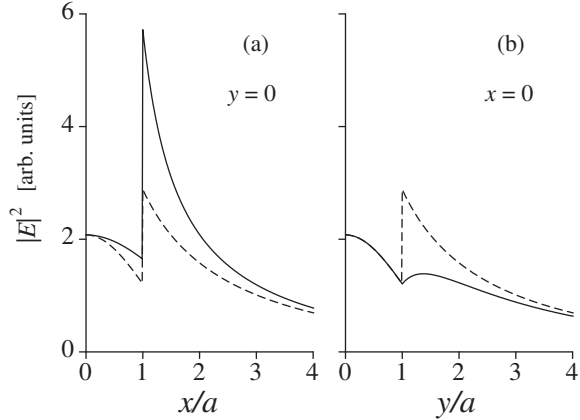}
 \end{center}
\caption{Total intensity $|E|^2$ of the electric field in a fundamental mode with  quasi-linear polarization
as a function of $x$ at $y=0$ (a) and as a function of $y$ at $x=0$ (b). 
For comparison, the intensity of the electric field in the corresponding approximate mode LP$_{01}$ is shown
by the dashed lines.
The parameters  for this figure  are the same as those for Fig.~\ref{fig2}.}
\label{fig3}
\end{figure}

To get a look at the spatial distributions of the field   along different radial  directions,
we replot in Fig.~\ref{fig3} the intensity $|E|^2$ of Fig.~\ref{fig2} 
as a function of $x$ at $y=0$ (as a function of $r$ at $\varphi=0$)  
and as a function of $y$ at $x=0$ (as a function of $r$ at $\varphi=\pi/2$).
For comparison, we plot by the dashed lines
the intensity $|E_{\mathrm{LP}}|^2$ of the corresponding  approximate mode LP$_{01}$, which is given by
$|E_{\mathrm{LP}}|^2=g_{\mathrm{in}}J_0^2(hr)$ for $r<a$ and 
$|E_{\mathrm{LP}}|^2=g_{\mathrm{out}}K_0^2(qr)$ for $r>a$. The comparison between  the region $x/a>1$ of Fig.~\ref{fig3}(a) and the region $y/a>1$ of Fig.~\ref{fig3}(b) shows that, outside the fiber, $|E|^2$ decays along the $x$ direction  faster than  along the $y$ direction. We observe that
the radial dependence of $|E|^2$ along the $y$ direction has a small peak in the outer vicinity of the fiber surface. Far away from the fiber surface, the decays of $|E|^2$ along the $x$ and $y$ directions
tend to have the same behavior as that of the intensity $|E_{\mathrm{LP}}|^2$ of the approximate mode LP$_{01}$. 
The comparison between  the region $x/a<1$ of Fig.~\ref{fig3}(a) and the region $y/a<1$ of Fig.~\ref{fig3}(b) shows that, inside the fiber, $|E|^2$ decreases along the $x$ direction  slower than  along the $y$ direction.
A discontinuity of $|E|^2$ is observed at the fiber surface in the $x$ direction, but not in the $y$ direction.

\begin{figure}
\begin{center}
  \includegraphics{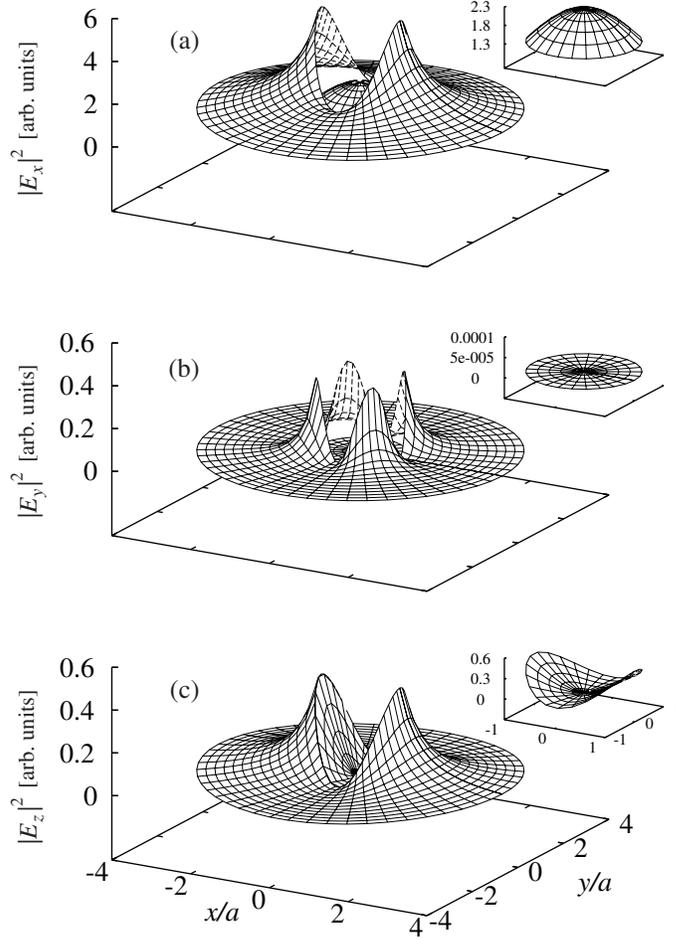}
 \end{center}
\caption{Cross-section profiles of the  intensities $|E_x|^2$, $|E_y|^2$, and $|E_z|^2$ of the 
Cartesian-coordinate  components of the electric field in a fundamental mode with  quasi-linear polarization. 
The insets show the inner parts of the profiles, which correspond to the field inside  the fiber.
The parameters  for this figure  are the same as those for Fig.~\ref{fig2}.}
\label{fig4}
\end{figure}

\begin{figure}
\begin{center}
  \includegraphics{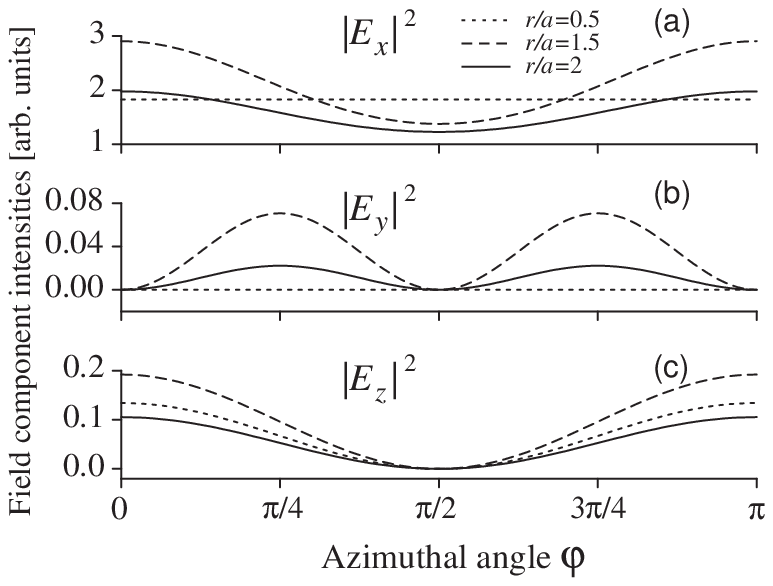}
 \end{center}
\caption{Azimuthal profiles  of the  intensities $|E_x|^2$, $|E_y|^2$, and $|E_z|^2$ of the Cartesian-coordinate  components of the electric field in a fundamental mode  with quasi-linear polarization.
The parameters  for this figure  are the same as those for Fig.~\ref{fig2}.}
\label{fig5}
\end{figure}

To get insight into the field components of the fundamental mode, we plot  in Fig.~\ref{fig4}
the cross-section profiles of the intensities $|E_x|^2$, $|E_y|^2$, and $|E_z|^2$ of the 
Cartesian-coordinate  components of the electric field. 
In addition, we plot in Fig.~\ref{fig5} these intensities as functions of the azimuthal angle $\varphi$.
The comparison between the scales of the vertical axes of  Figs.~\ref{fig4}(a), \ref{fig4}(b), and \ref{fig4}(c) shows that the intensity $|E_y|^2$ of the minor transverse component $E_y$ and  the intensity $|E_z|^2$ 
of the longitudinal component $E_z$ are  weaker  than  the intensity $|E_x|^2$ of the major transverse component $E_x$. However, $|E_y|^2$ and $|E_z|^2$ are not negligible at all. Therefore, the alignment of the total electric field vector $\mathbf{E}=E_x\hat{\mathbf{x}}+E_y\hat{\mathbf{y}}+E_z\hat{\mathbf{z}}$ may be substantially deviated from the major orientation axis $x$ of polarization. 
According to Eqs.~(\ref{5}) and (\ref{7}), the major and minor transverse components $E_x$ and $E_y$ of the electric field have the same phase. Therefore, the vector orientation of the total transverse component  $\mathbf{E}_\perp=E_x\hat{\mathbf{x}}+E_y\hat{\mathbf{y}}$ does not vary in time. 
Thus the transverse component $\mathbf{E}_\perp$ of the field is linearly polarized with respect to the time evolution at each fixed spatial point $(x,y)$. Meanwhile, the phase of the longitudinal component $E_z$ differs from that of the  transverse components $E_x$ and $E_y$ by $\pi/2$. Due to this phase difference, 
the total electric field  $\mathbf{E}$ rotates elliptically with time, in a plane parallel to the fiber axis $z$.

The inset in Fig.~\ref{fig4}(a) and the dotted curve in Fig.~\ref{fig5}(a) show that, inside  the fiber,  $|E_x|^2$ practically does not vary with $\varphi$. The inset in Fig.~\ref{fig4}(b) and the dotted curve in Fig.~\ref{fig5}(b) show that, inside  the fiber, $|E_y|^2$ is negligibly small, so the orientation of the vector $\mathbf{E}_\perp$ almost coincides with the $x$ axis. Thus the total transverse component $\mathbf{E}_\perp$ is not only linearly polarized with respect to the time evolution at each fixed local point but also is almost linearly polarized in the space inside the fiber. 

The main profiles in Figs.~\ref{fig4}(a) and \ref{fig4}(b) as well as
the dashed and solid curves in Figs.~\ref{fig5}(a) and \ref{fig5}(b) show that, outside the fiber, $|E_x|^2$ and $|E_y|^2$  substantially depend on $\varphi$. This behavior is quite different from the behavior of the field inside  the fiber. 

Figures \ref{fig4}(c) and \ref{fig5}(c) show that the longitudinal-component intensity $|E_z|^2$ is substantial  not only in the outer vicinity of the fiber surface but also in the inner vicinity, 
unlike the minor-transverse-component intensity $|E_y|^2$. In addition, 
$|E_z|^2$ substantially varies with $\varphi$ not only outside  but also inside the fiber,
unlike the major-transverse-component intensity $|E_x|^2$.
The azimuthal dependence of $|E_z|^2$ is profound in both sides of the fiber surface because
$E_z$ is proportional to $\cos(\varphi-\varphi_0)$, see  Eqs.~(\ref{5}) and (\ref{7}).

\begin{figure}
\begin{center}
  \includegraphics{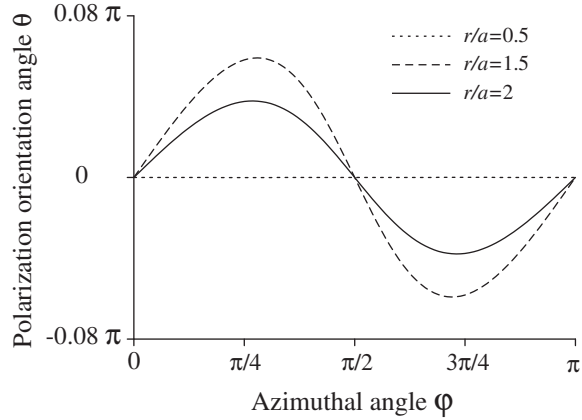}
 \end{center}
\caption{Orientation angle $\theta=\arctan[\mathrm{Re\,}(E_y)/\mathrm{Re\,}(E_x)]$ of the transverse component
$\mathbf{E}_\perp$ of the electric field in  a fundamental mode with  quasi-linear polarization.
The parameters  for this figure  are the same as those for Fig.~\ref{fig2}.}
\label{fig6}
\end{figure}

At each fixed spatial point, the vector orientation of the total transverse component $\mathbf{E}_\perp$ of the electric field does not, as discussed earlier, vary  in time. However, the orientation of $\mathbf{E}_\perp$ may  vary in space, especially in the outer vicinity of the fiber surface, where the minor transverse component $E_y$ is substantial. We plot in Fig.~\ref{fig6} the  orientation angle $\theta=\arctan[\mathrm{Re\,}(E_y)/\mathrm{Re\,}(E_x)]$ of $\mathbf{E}_\perp$ as a function of $\varphi$. The dotted curve in Fig.~\ref{fig6} shows that, inside  the fiber, the orientation angle $\theta$ almost does not vary with $\varphi$. This feature confirms that the total transverse component $\mathbf{E}_\perp$ inside the fiber is almost linearly polarized in  space. It is due to the fact that the $\varphi$-dependent terms in the expressions for the transverse components inside the fiber  are negligible. Meanwhile, the dashed and solid curves in Fig.~\ref{fig6}  show that, outside the fiber, the orientation angle $\theta$ does vary in space. 
For $r/a=1.5$, the maximal azimuthal variation of $\theta$   is about 0.06 $\pi$ (see the dashed curve). 
The comparison between the solid and dashed curves in Fig.~\ref{fig6} shows that, when $r$  increases from $a$, the azimuthal variation of $\theta$  reduces. These features are due to the fact that the ratio between the factors $K_2(qr)$ and $K_0(qr)$ of the $\varphi$-dependent and -independent terms is large for $qr\ll1$ but  monotonically reduces to unity with increasing $r$.

\section{Fundamental mode with rotating polarization}
\label{sec:circular}

We now study a fundamental mode  with rotating (circulating) polarization.
In the cylindrical coordinates, the solutions of  Maxwell's equations for 
the cylindrical components of the electric field $\mathbf{E}$ in such a mode are given by \cite{fiber books}
\begin{eqnarray}
E_r&=&F_r e^{\pm i\varphi}\,e^{i(\omega t-\beta z)},
\nonumber\\
E_\varphi&=&\pm F_\varphi e^{\pm i\varphi}\,e^{i(\omega t-\beta z)},
\nonumber\\
E_z&=& F_z e^{\pm i\varphi}\,e^{i(\omega t-\beta z)}.
\label{1b}
\end{eqnarray} 
Here the functions $F_j$ ($j=r,\varphi,z$) describe the radial dependences of the field components.
They are defined, for $r<a$, as
\begin{eqnarray}
F_r&=&-iA\frac{\beta}{2h} [(1-s)J_0(hr)-(1+s)J_2(hr) ],
\nonumber\\
F_\varphi&=&A\frac{\beta}{2h} [(1-s)J_0(hr)+(1+s)J_2(hr) ],
\nonumber\\
F_z&=& AJ_1(hr),
\label{1}
\end{eqnarray}
and, for $r>a$, as
\begin{eqnarray}
F_r&=&-iA\frac{\beta}{2q}\frac{J_1(ha)}{K_1(qa)} [(1-s)K_0(qr)+(1+s)K_2(qr) ],
\nonumber\\
F_\varphi&=&A\frac{\beta}{2q}\frac{J_1(ha)}{K_1(qa)} [(1-s)K_0(qr)-(1+s)K_2(qr) ],
\nonumber\\
F_z&=& A\frac{J_1(ha)}{K_1(qa)}K_1(qr).
\label{2}
\end{eqnarray} 
The upper (lower) sign in Eqs.~(\ref{1b}) corresponds to the clockwise (counterclockwise) circulation of photons around the $z$ axis.
According to Eqs.~(\ref{1b}), we have $|E_j|=|F_j|$. According to Eqs.~(\ref{1}) and (\ref{2}), the functions $F_j$ are independent of the azimuthal angle $\varphi$. Hence, the intensities $|E_j|^2$  of the cylindrical-coordinate components  of the field are independent of $\varphi$, and so is the total  intensity $|E|^2$  of the electric  field.

According to Eqs.~(\ref{1}) and (\ref{2}),  we have 
$F_r/|F_r|=-iF_\varphi/|F_\varphi|$. Then, it follows from Eqs.~(\ref{1b}) that 
$E_r/|E_r|=\mp iE_\varphi/|E_\varphi|$, that is, the two complex quadratures $E_r$ and $E_\varphi$ have a difference of $\pi/2$ between their phases. Therefore, the polarization of the total transverse component $\mathbf{E}_\perp=E_r\hat{\mathbf{r}}+E_\varphi\hat{\mbox{\boldmath $\varphi$}}$ of the field is either circular or elliptical. With increasing time, the real part of the complex amplitude vector 
$\mathbf{E}_\perp$ rotates along a circle or an ellipse. The semimajor and semiminor axes of the ellipse are aligned along the unit vectors $\hat{\mathbf{r}}$ and $\hat{\mbox{\boldmath $\varphi$}}$,  respectively, and are equal to $|E_r|=|F_r|$ and $|E_\varphi|=|F_\varphi|$, respectively. When $|E_r|\cong|E_\varphi|$, the elliptical  polarization of the transverse component of the field becomes almost circular.

We can easily calculate  the total intensity $|E|^2$ of the electric field in a fundamental mode  with rotating polarization. For the  field inside the fiber, we obtain
\begin{equation}
|E|^2=2g_{\mathrm{in}}[J_0^2(hr)+ u J_1^2(hr)+fJ_2^2(hr)].
\label{10}
\end{equation}
For the field outside the fiber, we get
\begin{equation}
|E|^2=2g_{\mathrm{out}}[K_0^2(qr)+ w K_1^2(qr)+fK_2^2(qr)].
\label{9}
\end{equation}
The terms $J_0^2(hr)$ and  $K_0^2(qr)$ in the expressions (\ref{10}) and (\ref{9}), respectively, correspond to the total intensity of the electric field in the mode LP$_{01}$. 
The other terms  describe the deviations of the exact fundamental mode HE$_{11}$ with rotating polarization from the  approximate  mode LP$_{01}$. 
The comparison between Eqs.~(\ref{12}) and (\ref{10}) and
between Eqs.~(\ref{11}) and (\ref{9}) 
shows that the total intensity of the electric field in a fundamental mode with
rotating polarization is the sum of the corresponding intensities for two constituent modes with
quasi-linear polarizations. The $\varphi$-dependent terms  cancel each other and therefore do not appear 
in Eqs.~(\ref{10}) and (\ref{9}).

The above expressions  are mathematically valid for the fundamental mode with rotating polarization of a fiber 
with an arbitrary core radius $a$ and an arbitrary pair of refractive indices $n_1>n_2$. 
To demonstrate the features of vacuum-clad \textit{subwavelength-diameter} fibers, we perform numerical calculations for the parameters of the previous section, namely, for 
$a=0.2$ $\mu$m, $\lambda=1.3$ $\mu$m, $n_1\cong 1.4469$, and $n_2=1$. 

\begin{figure}
\begin{center}
  \includegraphics{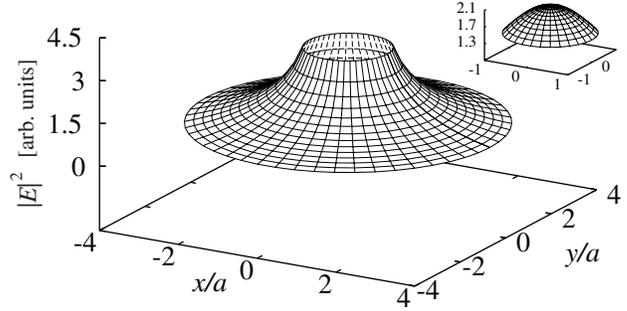}
 \end{center}
\caption{Cross-section profile of the total intensity $|E|^2$ of the electric field in a fundamental mode with  
rotating polarization. The inset shows the inner part of the profile, which corresponds to the field inside  the fiber. The parameters  for this figure  are the same as those for Fig.~\ref{fig2}.}
\label{fig7}
\end{figure}

In Fig.~\ref{fig7}, we plot the cross-section profile of the total intensity $|E|^2$ of the electric field in a  fundamental mode with rotating polarization. As seen, $|E|^2$ is azimuthally independent, that is, is cylindrically symmetric.
The discontinuity of the field at the fiber surface, created by the boundary condition and the high contrast between $n_1$ and $n_2$, divides the profile into two parts. 
The inner part of the profile  is hidden behind the outer part and is shown separately in the inset of the figure.

\begin{figure}
\begin{center}
  \includegraphics{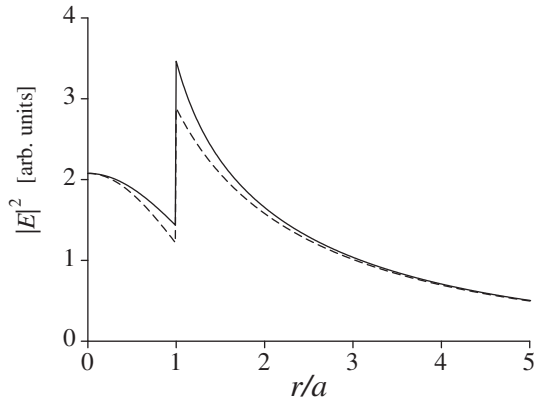}
 \end{center}
\caption{Radial dependence of the total intensity $|E|^2$ of the electric field in a fundamental mode with  rotating polarization (solid line). For comparison, 
the intensity $|E_{\mathrm{LP}}|^2$, 
obtained for the corresponding approximate mode LP$_{01}$,
is plotted by the dashed line. 
The parameters  for this figure  are the same as those for Fig.~\ref{fig2}.}
\label{fig8}
\end{figure}

In Fig.~\ref{fig8}, we plot by the solid line the total intensity $|E|^2$ of the electric field  as a function of the radial distance $r$. 
For comparison, we plot by the dashed line
the intensity $|E_{\mathrm{LP}}|^2$ of the corresponding approximate mode LP$_{01}$, which is given by
$|E_{\mathrm{LP}}|^2=2g_{\mathrm{in}}J_0^2(hr)$ for $r<a$ and 
$|E_{\mathrm{LP}}|^2=2g_{\mathrm{out}}K_0^2(qr)$ for $r>a$.
As seen, the difference between $|E|^2$ and $|E_{\mathrm{LP}}|^2$ is small but not negligible
in the region $0.3<r/a<3$. 
Inside the fiber, $|E|^2$ (solid curve) decreases slower than $|E_{\mathrm{LP}}|^2$ (dashed curve). 
Such a behavior of $|E|^2$ is  due to the contributions of the last two terms in Eq.~(\ref{10}).  
These terms contain the functions $J_1$ and $J_2$, which increase in the region of small argument.
In the outer vicinity of the fiber surface, $|E|^2$ (solid curve) decays faster than $|E_{\mathrm{LP}}|^2$ (dashed curve). Such a behavior of $|E|^2$ is  due to the contributions of the additional terms $wK_1^2(qr)$ and  $fK_2^2(qr)$ in Eq.~(\ref{9}), which  decay faster than the basic term $K_0^2(qr)$ in the region of small argument. When $r/a$ is large enough ($r/a>3$), the difference between $|E|^2$ and
$|E_{\mathrm{LP}}|^2$ can be neglected.

\begin{figure}
\begin{center}
  \includegraphics{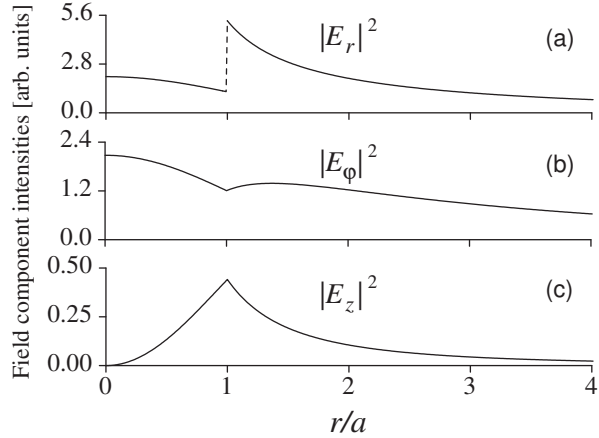}
 \end{center}
\caption{Intensities  $|E_r|^2$, $|E_\varphi|^2$, and $|E_z|^2$ of the cylindrical-coordinate components of the field in a fundamental mode with rotating polarization. 
The parameters  for this figure  are the same as those for Fig.~\ref{fig2}.}
\label{fig9}
\end{figure}

The intensities $|E_r|^2$, $|E_\varphi|^2$, and $|E_z|^2$
of the cylindrical-coordinate components of the field
are determined by the cylindrically symmetric functions $|F_r|^2$, $|F_\varphi|^2$, and $|F_z|^2$, respectively. 
We plot these intensities as functions of $r$ in Fig.~\ref{fig9}.
As seen,  the field intensity distributions  inside ($r/a<1$) and outside ($r/a>1$) the fiber have very different behaviors. Due to the boundary condition and the high  contrast between the refractive indices of the silica core and the vacuum clad, the normal (radial) component $E_r$ has a conspicuous discontinuity at the fiber surface. The tangential components $E_\varphi$ and $E_z$ are, however, continuous.

\begin{figure}
\begin{center}
  \includegraphics{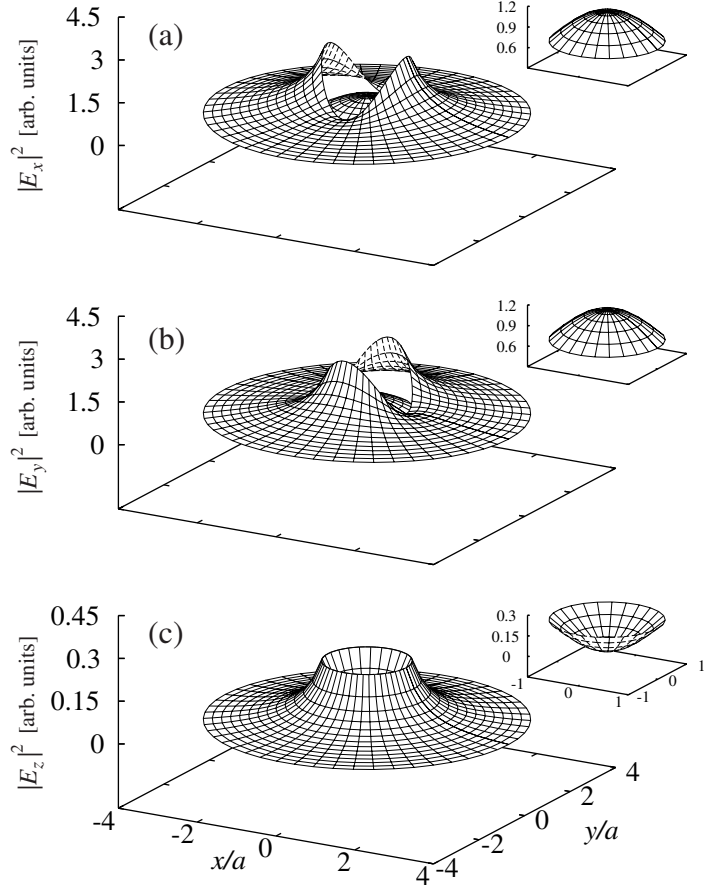}
 \end{center}
\caption{Cross-section profiles of the  intensities $|E_x|^2$, $|E_y|^2$, and $|E_z|^2$ of the 
Cartesian-coordinate components of the electric field in a fundamental mode with rotating polarization. 
The insets show the inner parts of the profiles, which correspond to the field inside  the fiber. The parameters  for this figure  are the same as those for Fig.~\ref{fig2}.}
\label{fig10}
\end{figure}

\begin{figure}
\begin{center}
  \includegraphics{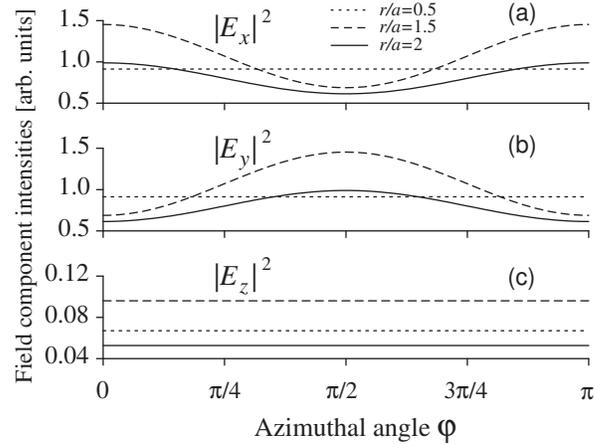}
 \end{center}
\caption{Azimuthal profiles  of the  intensities $|E_x|^2$, $|E_y|^2$, and $|E_z|^2$ of the 
Cartesian-coordinate components of the electric field in a fundamental mode with  rotating polarization. 
The parameters  for this figure  are the same as those for Fig.~\ref{fig2}.}
\label{fig11}
\end{figure}

Unlike the intensities of the cylindrical-coordinate components $E_r$, $E_\varphi$, and  $E_z$, the intensities of the Cartesian-coordinate transverse components $E_x$ and $E_y$ are, in general,  not cylindrically symmetric. 
In Fig.~\ref{fig10}, we plot the cross-section profiles of the  intensities $|E_x|^2$, $|E_y|^2$, and $|E_z|^2$. 
In addition, we plot these intensities in Fig.~\ref{fig11} as functions of $\varphi$.

The insets in Figs.~\ref{fig10}(a) and \ref{fig10}(b) as well as the dotted curves in Figs.~\ref{fig11}(a) and \ref{fig11}(b) show that, inside  the fiber, $|E_x|^2$ and $|E_y|^2$ are almost equal to each other and practically do not vary with $\varphi$. The reason is that the $\varphi$-dependent terms in the expressions
for $F_r$ and $F_\varphi$ in Eqs.~(\ref{1}) are negligible.  
Because of this, we have $F_r\cong -iF_\varphi$ and, consequently, $E_r\cong \mp iE_\varphi$. 
Hence, we find $E_x\cong F_re^{i(\omega t-\beta z)}$, $E_y\cong \pm i F_re^{i(\omega t-\beta z)}$,  and consequently $E_x\cong \mp iE_y$. 
Thus the two orthogonal quadratures  $E_r$ and $E_\varphi$ as well as $E_x$ and $E_y$ have almost equal magnitudes 
and a relative phase difference of $\pi/2$. 
This indicates that, inside the fiber, the total transverse component 
$\mathbf{E}_\perp$ of the field  is  almost circularly polarized.

The main profiles in Figs.~\ref{fig10}(a) and \ref{fig10}(b) as well as 
the dashed and solid curves in Figs.~\ref{fig11}(a) and \ref{fig11}(b)  
show that, outside the fiber, the intensities of the transverse components $E_x$ and $E_y$ are different from each other  and substantially vary with  $\varphi$. The comparison between the dashed lines and the solid lines in Figs.~\ref{fig11}(a) and \ref{fig11}(b) shows that the azimuthal dependences of $|E_x|^2$ and $|E_y|^2$ 
outside the fiber reduce with increasing $r$. 

The terms $\pm(1+s)K_2(qr)$ in Eqs.~(\ref{2}) lead to $|F_r|\not=|F_\varphi|$ and, hence,
to $|E_r|\not=|E_\varphi|$.
Due to the substantial difference between  $|E_r|$ and $|E_\varphi|$ in the outer vicinity of the fiber surface,
the polarization of the transverse component $\mathbf{E}_\perp$ of the field  in this region  is truly elliptical.
We recall that the axes of the elliptical circulation orbit are aligned along the unit vectors $\hat{\mathbf{r}}$ and $\hat{\mbox{\boldmath $\varphi$}}$. Consequently, 
the orientation of the orbit varies with increasing $\varphi$. We also recall that 
the lengths of the axes of the ellipse are equal to 
$|E_r|=|F_r|$ and $|E_\varphi|=|F_\varphi|$. Hence, the ellipticity of the orbit is proportional to the factor $(1+s)K_2(qr)/(1-s)K_0(qr)$. This factor reduces with increasing $r$. 
Thus both the orientation and the ellipticity of the orbit of polarization circulation  vary in space. 
This behavior is different from the  case of conventional light beams with  elliptical or  circular polarization.
Furthermore, we note that the ellipticity of the orbit of polarization circulation does not depend on $\varphi$.
This indicates that the orbit rotates circularly in space.

Figures \ref{fig10}(c) and \ref{fig11}(c) show that the longitudinal-component intensity $|E_z|^2$ is perfectly cylindrically symmetric, as expected, in the whole cross-section plane. Although $|E_z|^2$ is small compared to  $|E_x|^2$ and $|E_y|^2$, it is not negligible in the vicinity of the fiber surface.

\section{Conclusions}
\label{sec:summary}

We have studied the properties of the field in the fundamental mode HE$_{11}$ of a vacuum-clad \textit{subwavelength-diameter} optical fiber using the exact solutions of Maxwell's equations. 
We have obtained simple analytical  expressions for the total intensity of the electric field and have identified the origin of the deviations of the exact fundamental mode HE$_{11}$ from the approximate mode LP$_{01}$. 
We have shown that the thin thickness of the fiber and the high  contrast between the refractive indices of the silica core and the vacuum clad substantially modify the intensity distributions and the polarization properties of the field and its components, especially in the  vicinity of the fiber surface. 

We have  examined the case of a fundamental mode with quasi-linear polarization. 
We have  shown that a substantial azimuthal dependence of
the total intensity is observed in the  vicinity of the fiber surface.
The transverse component  of the field is linearly polarized in time at each fixed local point. 
However, the total electric field vector   rotates elliptically with time, in a plane parallel to the fiber axis. 
Inside the fiber, the transverse component is not only linearly polarized in time but also almost linearly polarized in space.
Outside the fiber, the orientation angle of the transverse component of the field  varies in space.

We have also studied the case of a fundamental mode with rotating (circulating) polarization. 
We have  shown that the total intensity  is azimuthally independent, that is, is cylindrically symmetric. 
We have found that the total intensity  of the electric field outside the fiber decays faster
than that of the approximate mode LP$_{01}$. The difference between the exact and approximate modes is relatively small in the case of rotating polarization. Consequently, the underlying physics
of the optical potential of the evanescent wave around a vacuum-clad subwavelength-diameter fiber is basically the same as that of the approximate mode LP$_{01}$. However, the magnitude of the difference is not negligible
in the vicinity of the fiber surface. Therefore, the use of the exact solutions of Maxwell's equations 
is required  in a systematic quantitative treatment for a thin fiber. 
We have shown that, inside the fiber, the total transverse component 
of the field  is  almost circularly polarized. In the outer vicinity of the fiber surface, 
the polarization of the transverse component  of the field  is elliptical.
In this region, the orientation of the orbit of polarization circulation rotates in space while 
the ellipticity of the orbit reduces with increasing radial distance. 
This is different from the  case of conventional light beams with  elliptical or  circular polarization.
Our results are helpful for studying and developing new miniaturized high-performance photonic devices.
One of the promising applications of the field around the subwavelength-diameter fiber is  trapping and guiding of neutral atoms by the optical force of the evanescent field. 

\section*{Acknowledgments}

We thank V. V. Klimov for  prompting us to start this work by giving
a comment about the cylindrical asymmetry of the field in the fundamental mode of a thin fiber.
This work was carried out under the 21st Century COE program on ``Coherent Optical Science''.

\end{document}